\documentclass[aps,amssymb,prd,showpacs,twocolumn]{revtex4}
\usepackage{amsmath}
\begin{document}

\title{Einstein boundary conditions for the Einstein
equations in the conformal-traceless decomposition}

\author{Simonetta Frittelli}
\email{simo@mayu.physics.duq.edu}
\affiliation{Department of Physics, Duquesne University,
       Pittsburgh, PA 15282}
\affiliation{Department of Physics and Astronomy, University of Pittsburgh,
       Pittsburgh, PA 15260}
\author{Roberto G\'omez}
\email{gomez@psc.edu}
\affiliation{Pittsburgh Supercomputing Center,
Carnegie Mellon University, Pittsburgh, PA 15213,}
\affiliation{Department of Physics and Astronomy, University of Pittsburgh,
       Pittsburgh, PA 15260}

\date{\today}

\begin{abstract}

In relation to the BSSN formulation of the Einstein equations, we write down
the boundary conditions that result from the vanishing of the projection of the
Einstein tensor normally to a timelike hypersurface. Furthermore, by setting
up a well-posed system of propagation equations for the constraints, we show
explicitly that there are three constraints that are incoming at the boundary
surface and that the boundary equations are linearly related to them. This
indicates that such boundary conditions play a role in enforcing the
propagation of the constraints in the region interior to the boundary.
Additionally, we examine the related problem for a strongly hyperbolic
first-order reduction of the BSSN equations and determine the characteristic
fields that are prescribed by the three boundary conditions, as well as those
that are left arbitrary. 

\end{abstract}
\pacs{04.25.Dm, 04.20.Ex}
\maketitle

%-------------------------------------------------------------------
\section{Introduction}\label{sec:1}

In a series of papers~\cite{0302032,boundary3d,bcconst} we have carried out a
study of the relevance of the projection of the Einstein tensor $G_{ab}$
perpendicularly to a timelike surface as candidate boundary conditions for the
initial-boundary value problem of the Einstein equations.  Using the
Einstein-Christoffel formulation~\cite{fixing} we demonstrated that the
vanishing of three of the four components of the projection $G_{ab}e^b$ (where
$e^b$ is the vector normal to the timelike boundary) are the three necessary
and sufficient boundary conditions required for the initial value problem, and
that they are also equivalent to the three necessary and sufficient boundary
conditions required for the well-posed propagation of the vanishing of the
constraints.  Because of the generality of the methods used to demonstrate this
result, we believe that such will be the case for any formulation of the
3+1 Einstein equations that is strongly hyperbolic.  

The full significance of $G_{ab}e^b=0$ to the more standard 3+1 formulation
referred to as ADM~\cite{yorksources} is not as well established, essentially
due to the lack of hyperbolicity of the ADM equations themselves. Yet one can
see that failure to impose at least one of the four equations leads to
constraint violations in the solution at later times~\cite{bcconst}. 

At this time, we are intersted in investigating the relevance of the Einstein
boundary conditions $G_{ab}e^b=0$ to the ``conformal-traceless'' decomposition
of the 3+1 equations due to Baumgarte and Shapiro~\cite{BSSN}, which is
developed on the basis of earlier work by Shibata and Nakamura~\cite{SN}, and
is known as BSSN.  We have two reasons for interest in this formulation. In
the first place, it is a hybrid between the ADM equations and the more recent
hyperbolic formulations, because it represents a partial reduction to
first order of the original ADM equations with mixing of the constraints in a
manner similar to most hyperbolic formulations (see, e.g., \cite{simo96}).  As
such, it represents an opportunity to investigate the transition between ADM
and its hyperbolic (well-posed) reductions. 

Our other reason for interest stems from the fact that numerical simulations
performed using codes based on the BSSN formulation appear to be able to run for significantly long
times, which makes it, currently, a formulation of widespread choice in
numerical relativity (see, e.g., \cite{laguna}), up to slight modifications
(see, e.g.,~\cite{bruegmann,laguna2}). 

In Section~\ref{sec:2} we briefly describe the problem of the Einstein boundary
conditions for the ADM equations with the purpose of introducing the Einstein
boundary conditions in terms of the fundamental variables of the 3+1 split. The
Einstein boundary conditions in terms of the BSSN variables are introduced in
Section~\ref{sec:3}, whereas their relationship to the propagation of the
constraints is developed in Section~\ref{sec:4}.  In Section~\ref{sec:5} we
study the analogous problem for the case of a well-posed first-order reduction
of the BSSN equations. Our first-order reduction is strongly hyperbolic, having
six characteristic fields that are incoming at the boundary, and six that are
outgoing at the boundary. Unlike the Einstein-Christoffel formulation and other
first-order reductions of the BSSN equations, it has two subluminal
characteristic speeds in addition to light speed and rest. We show explicitly
how the Einstein boundary conditions prescribe only three of the incoming
characteristic fields, which is entirely consistent with the results of
Section~\ref{sec:4}, and coincides with our previous results on the
Einstein-Christoffel formulation.   As in our previous work, we restrict to the case of
vanishing shift vector throughout.  We close in Section~\ref{sec:6} with some
remarks.

%-------------------------------------------------------------------
\section{The Einstein boundary conditions in terms of the intrinsic and
extrinsic curvature for the case of vanishing shift vector\label{sec:2}}

This section develops a brief review of the boundary conditions of
the Einstein equations in the ADM formulation.  In our minds, the
ADM case acts as the basis for most other formulations that are
derived from it by linear combinations with the constraints and by
the addition of any number of first-order variables.

Throughout the article we assume the following form for the metric of spacetime
in coordinates $x^a = (x^i,t)$ in terms of the three-metric $\gamma_{ij}$ of
the slices at fixed value of $t$:
\begin{equation}\label{metric}
    ds^2 = -\alpha^2 dt^2 + \gamma_{ij}dx^idx^j
\end{equation}

\noindent where $\alpha$ is the lapse function. The Einstein equations
$G_{ab}=0$ for the four-dimensional metric are equivalently expressed in the
ADM form~\cite{yorksources}:
\begin{subequations}\label{adm}
\begin{eqnarray}
    \dot{\gamma}_{ij} &=& - 2\alpha K_{ij}, \label{adma}\\
    \dot{K}_{ij} &=& \alpha \left(R_{ij} - 2 K_{il}K^l{}_j
            +K K_{ij}\right) -D_iD_j\alpha, \label{admb}
\end{eqnarray}
\end{subequations}

\noindent with the constraints
\begin{subequations}\label{admconst}
\begin{eqnarray}
    {\cal C} &\equiv &\frac12\left(R - K_{ij}K^{ij} + K^2 \right)= 0,   \\
    {\cal C}_i&\equiv &D_jK^j{}_i-D_i K = 0.
\end{eqnarray}
\end{subequations}

\noindent Here an overdot denotes a partial derivative with
respect to the time coordinate ($\partial/\partial t$), indices
are raised with the inverse metric $\gamma^{ij}$, $D_i$ is the
covariant three-derivative consistent with $\gamma_{ij}$ ,
$R_{ij}$ is the Ricci curvature tensor of $\gamma_{ij}$, $R$ its
Ricci scalar, $K_{ij}$ is the extrinsic curvature of the slice at
fixed value of $t$ and $K\equiv \gamma^{ij}K_{ij}$.  Expressed in
terms of the Einstein tensor, the constraints (\ref{admconst}) are
related to specific components in the coordinates $(x^i,t)$:
\begin{subequations}\label{constG}
\begin{eqnarray}
    {\cal C}  &=& \alpha^2 G^{tt}\label{constGa}\\
    {\cal C}_i&=& -\alpha \gamma_{ij}G^{jt} ,\label{constGb}
\end{eqnarray}
\end{subequations}

\noindent where (\ref{constGb}) holds only for vanishing shift
vector. The constraint character (the absence of second
derivatives with respect to time) is a consequence of the fact
that the only components of the Einstein tensor that appear in
(\ref{constG}) have a contravariant index of value $t$. In
geometric terms, (\ref{constG}) are linear combinations of
$G_{ab}n^b=0$ where $n^b$ is the unit normal to the slices of
fixed value of $t$, and is therefore given by
$n^a=g^{ab}n_b=-\alpha g^{at}=\delta^a_t/\alpha$.

Similarly, at any boundary given by a fixed value of a spatial
coordinate, the normal vector to the boundary surface $e^b$ can be
used to project the Einstein tensor as $G_{ab}e^b$ in order to
obtain the components that have no second derivatives across the
boundary. As we have shown previously~\cite{boundary3d}, the
vanishing of these can then be imposed as conditions on the
boundary values for the fundamental variables. To fix ideas let's
choose a boundary surface at a fixed value of $x$. Thus $e^b=
g^{bx}= (0,\gamma^{ix})$ up to scaling. Consequently, $G_{ab}e^b =
G_a^x$, so any linear combination of the components of the
Einstein tensor with a contravariant index of value $x$ will be
suitable. Explicitly we have:
\begin{subequations}\label{bcs}
\begin{eqnarray}
G_t^x &=& -\frac12 \gamma^{ix}\left((\ln \gamma),_{it}
          -\gamma^{kl}\dot{\gamma}_{ik,l}\right)
-KD^x\alpha +K^x_kD^k\alpha\nonumber\\ &&
         + \alpha\left( \gamma^{kl}\Gamma^j{}_{kl}K^x_j
           +\gamma^{ix}\Gamma^j{}_{ik}K^k_j\right) \label{det}
\\
%\end{eqnarray}
%\begin{eqnarray}
G_x^x&=& \frac{\dot{K}-\dot{K}^x_x}{\alpha}
-\frac12 (R +K^{ij}K_{ij} +K^2) +KK^x_x\nonumber\\
&&
         +R^x_x+\frac{1}{\alpha}\left(D^jD_j\alpha- D^xD_x\alpha\right)
\label{K-Kxx}
%\\
\end{eqnarray}
\begin{eqnarray}
G_y^x &=& -\frac{\dot{K}^x_y}{\alpha} + KK^x_y
          + R^x_y -\frac{1}{\alpha}
        D^xD_y\alpha \label{Kxy}
 \\
%\end{eqnarray}
%\begin{eqnarray}
G_z^x &=& -\frac{\dot{K}^x_z}{\alpha} + KK^x_z
          + R^x_z -\frac{1}{\alpha}
        D^xD_z\alpha\label{Kxz}
\end{eqnarray}
\end{subequations}

\noindent  Here
$\Gamma^k{}_{ij}=(1/2)\gamma^{kl}(\gamma_{il,j}+\gamma_{jl,i}-\gamma_{ij,l})$,
and the time derivative of the components of the extrinsic
curvature is applied after raising an index, that is: $\dot{K}^i_j
\equiv (\gamma^{ik}K_{kj}),_t$. The reader can verify that $R^x_y$
and $R^x_z$ do not involve second derivatives with respect to $x$
of any of the variables and that the combination $R^x_x -\frac12
R$ doesn't either~\cite{0302032}.

Since Eqs.~(\ref{bcs}) represent the vanishing of four independent
components of the Einstein tensor, they must be expressible in
terms of linear combinations of the evolution equations
(\ref{adm}) and the constraints (\ref{admconst}). For this reason,
it could mistakenly be thought that they must be satisfied
identically by the solution of the evolution equations with
initial data satisfying the constraints.  Such would be the case
only in regions of spacetime where the constraints are satisfied,
as well as the evolution equations. Since the constraints are only
imposed on the initial data, it is not at all clear that they are
satisfied everywhere. A brief calculation
that involves taking a time derivative of the constraints and
using the evolution equations yields
\begin{subequations}\label{constprop}
\begin{eqnarray}
\dot{\cal C}  &=& \alpha \partial^i {\cal C}_i + \ldots \\
\dot{\cal C}_i &=& \alpha \partial_i{\cal C} + \ldots
\end{eqnarray}
\end{subequations}

\noindent where $\ldots$ denote undifferentiated terms. This system is 
strongly hyperbolic~\cite{kreissbook} because it has real characteristic speeds
and a complete set characteristic fields. With respect to the unit vector
$\xi^i = \gamma^{ix}/\sqrt{\gamma^{xx}}$, normal to the boundary, the
constraints ${\cal C}^y$ and ${\cal C}^z$ travel with zero characteristic
speeds, whereas ${\cal C}^\pm \equiv {\cal C} \pm {\cal
C}^x/\sqrt{\gamma^{xx}}$ travel with speeds $\pm\alpha$, respectively. Only
${\cal C}^y, {\cal C}^z$ and ${\cal C} + {\cal C}^x/\sqrt{\gamma^{xx}}$ are
vanishing at the boundary if they are set to zero on the region of the initial
slice interior to the boundary. Because ${\cal C} - {\cal
C}^x/\sqrt{\gamma^{xx}}$ is incoming, it is not vanishing at the boundary be
virtue of the evolution and the initial data alone, therefore it is an example
of what we refer to, colloquially, as an ``un-preserved'' constraint, and must
be made to vanish at the boundary by an appropriate choice of boundary
conditions. 

One may not, thus, dismiss the boundary equations (\ref{bcs}) as a matter of
principle. On the contrary, it is enlightening to see exactly how the boundary
equations relate to the constraints.

To start with, by inspection one can clearly see that $G_y^x$ and $G_z^x$ are
identical to the evolution equations (\ref{admconst}) for the mixed components
$K_y^x$ and $K_z^x$ of the extrinsic curvature (namely: the components $j=y$
and $j=z$ of Eq.~(\ref{admb}) multiplied by $\gamma^{ix}$). Thus two of the
four boundary equations are identically satisfied by the solution of the
evolution equations (irrespective of the initial data).  That is not the case
with $G_x^x$. If one uses the evolution equations for $K_x^x$ and for the trace
$K$ that follow from (\ref{admb}) to eliminate the time derivatives from
$G_x^x$, one is left with the Hamiltonian constraint ${\cal C}$.  Finally,
using the definition of the extrinsic curvature (\ref{adma}) in the
expression for $G_t^x$ it is straightforward to see that $G_t^x= \alpha
\gamma^{xi}{\cal C}_i$. Summarizing, if we represent  the evolution equations
(\ref{adma}) and (\ref{admb}) in the form ${\cal E}^\gamma_{ij}=0$ and ${\cal
E}^K_{ij}=0$ respectively by transferring all the terms from the right into the
left-hand side, we have
\begin{subequations}\label{bctoconst}
\begin{eqnarray}
G_t^x &=& \alpha{\cal C}^x
    + \frac12\gamma^{xj}\gamma^{kl}(\partial_l{\cal E}^{\gamma}_{jk}
    - \gamma^{xl}\gamma^{jk}\partial_l{\cal E}^{\gamma}_{jk})\label{bctoconsta}\\
G_y^x &= & \frac{1}{\alpha}\gamma^{xj}{\cal E}^{K}_{jy} \label{bctoconstb}\\
G_z^x &=& \frac{1}{\alpha}\gamma^{xj}{\cal E}^{K}_{jz}\label{bctoconstc}\\
G_x^x &= & {\cal C} +\frac{1}{\alpha}(\gamma^{kl}{\cal E}^K_{kl} - \gamma^{xj}{\cal
E}^K_{xj})\label{bctoconstd}
\end{eqnarray}
\end{subequations}

\noindent This means that $G_t^x$ and $G_x^x$ are equivalent to ${\cal C}^x$
and ${\cal C}$ respectively (modulo the evolution), whereas $G_y^x$ and $G_z^x$
are equivalent to zero (modulo the evolution).  One can see that imposing
either $G_t^x=0$ or $G_x^x=0$ is equivalent to a boundary condition of the form
${\cal C}^-= B{\cal C}^+$ with a constant $B$ on the system of evolution of
the constraints. This is a valid boundary condition in the sense that it
preserves the well-posedness of the system of evolution of the
constraints~\cite{kreissbook}. Thus, of the two components  $G_t^x=0$ or
$G_x^x=0$, one represents a necessary boundary condition. The other one (or a
linear combination of $G_t^x=0$ and  $G_x^x=0$ not equivalent to the first one)
becomes redundant because it represents a condition on the outgoing
characteristic field ${\cal C}^+$, which is already determined at the boundary
by the initial values. 

Because the ADM evolution equations are not strongly hyperbolic, perhaps
nothing else can be said about the role of the equations $G_{ab}e^b=0$ as
boundary conditions for the ADM equations, except that failure to impose the
nontrivial one leads to constraint violations with guaranteed certainty.

In the next Section, we examine the Einstein boundary conditions as applied in
the formulation of the Einstein equations that appears in ~\cite{BSSN}.

%-------------------------------------------------------------------
\section{Boundary conditions for the BSSN formulation with vanishing shift
\label{sec:3}}

In the case of vanishing shift vector, the formulation of the
Einstein equations referred to as BSSN~\cite{BSSN} consists of the
following evolution equations
\begin{subequations}\label{bssn}
\begin{eqnarray}
\dot{\tilde{\gamma}}_{ij} &=& -2\alpha \tilde{A}_{ij} \label{bssndotg}\\
\dot{\phi} &=& -\frac{\alpha}{6} K \label{bssndotphi}\\
%\dot{K} &=& \gamma^{ij}D_iD_j\alpha + \alpha(\tilde{A}_{ij}\tilde{A}^{ij}
%    +\frac13K^2)			\label{bssndotK}\\
%\dot{\tilde{\Gamma}}^i &=&
%    2\alpha\Big(\tilde{\Gamma}^i{}_{jk}\tilde{A}^{kj}
%+\frac23\tilde{\gamma}^{ij}K,_j
%     + 6 \tilde{A}^{ij}\phi,_j\Big)
%\nonumber\\ &&
%-2\tilde{A}^{ij}\alpha,_j		\label{dotGamma}
%\end{eqnarray}
%\begin{eqnarray}
\dot{\tilde{A}}_{ij} &=&\alpha e^{-4\phi} \bigg(-\frac12 \tilde{\gamma}^{lm}\tilde{\gamma}_{ij,lm}
    +\tilde{\gamma}_{k(i}\tilde{\Gamma}^k,_{j)}
    + \tilde{\Gamma}^k\tilde{\Gamma}_{(ij)k} \nonumber\\
&&
    +2\tilde{\Gamma}^{kl}{}_{(i}\tilde{\Gamma}_{i)kl}
    +\tilde{\Gamma}^{kl}{}_i\tilde{\Gamma}_{klj}
    -2\tilde{D}_i\tilde{D}_j\phi
    +4\tilde{D}_i\phi\tilde{D}_j\phi\nonumber\\
&&
    -\frac13\tilde{\gamma}_{ij}\big(\tilde{\Gamma}^k,_k
    +\tilde{\Gamma}^{kli}(2\tilde{\Gamma}_{ikl}
                  +\tilde{\Gamma}_{kli})\big)\nonumber\\
&&
    +\frac23\tilde{\gamma}_{ij}(\tilde{D}^l\tilde{D}_l\phi
    -2\tilde{D}^l\phi\tilde{D}_l\phi)
-{\frac{(D_iD_j\alpha)}{\alpha}}^{TF}\bigg)\nonumber\\
&&
            +K\tilde{A}_{ij}
-2\tilde{A}_{il}\tilde{A}^l_j \label{bssndotA}
\\
%\end{eqnarray}
%\begin{eqnarray}
\dot{K} &=& \gamma^{ij}D_iD_j\alpha + \alpha(\tilde{A}_{ij}\tilde{A}^{ij}
    +\frac13K^2)			\label{bssndotK}\\
\dot{\tilde{\Gamma}}^i &=&
    2\alpha\Big(\tilde{\Gamma}^i{}_{jk}\tilde{A}^{kj}
-\frac23\tilde{\gamma}^{ij}K,_j
     + 6 \tilde{A}^{ij}\phi,_j\Big)
\nonumber\\ &&
-2\tilde{A}^{ij}\alpha,_j		\label{dotGamma}
\end{eqnarray}
\end{subequations}

\noindent for the 15 variables
\begin{subequations}
\begin{eqnarray}
\phi&\equiv& \frac{1}{12}\ln (\det \gamma_{ij})\\
\tilde{\gamma}_{ij} &\equiv& e^{-4\phi}\gamma_{ij}\\
K &\equiv& \gamma^{ij}K_{ij}\\
\tilde{A}_{ij} &\equiv& e^{-4\phi}\left(K_{ij}
        -\frac13\tilde{\gamma}_{ij}K\right)\\
\tilde{\Gamma}^i &\equiv& -\tilde{\gamma}^{ij},_j
\end{eqnarray}
\end{subequations}

\noindent which mainly separate out the determinant of the
three-metric and the trace of the extrinsic curvature in order to
evolve them as fundamental variables in their own right. But
additionally, three first order variables are introduced,
$\tilde{\Gamma}^i$, corresponding to certain combinations of first
derivatives of the metric. This has the effect of modifying the
principal symbol of the system of PDE's.

For this system, we will express the boundary equations
(\ref{bcs}) directly in terms of the fundamental variables, in
such a way that no second $x-$derivatives of $\tilde{\gamma}_{ij}$
or $\phi$ occur, nor first $x-$derivatives of $\tilde{\Gamma}^i,
\tilde{A}_{ij}$ nor $K$ occur.

We start with the simplest one, that is $G^x_y$, which contains
second derivatives in the term $R^x_y$.

To start with, we have $R_{ij} = \tilde{R}_{ij}+R^\phi_{ij}$ where
$\tilde{R}_{ij}$ is the Ricci tensor of $\tilde{\gamma}_{ij}$ and
\begin{eqnarray}
R^\phi_{ij} &=& -2\tilde{D}_i\tilde{D}_j\phi
    -2\tilde{\gamma}_{ij}\tilde{D}^l\tilde{D}_l\phi\nonumber\\
&&
    +4\tilde{D}_i\phi\tilde{D}_j\phi
    -4\tilde{\gamma}_{ij}\tilde{D}^l\phi\tilde{D}_l\phi
\end{eqnarray}

\noindent So $R^x_y = \gamma^{xj}R_{yj} =
e^{-4\phi}\tilde{\gamma}^{xj}(\tilde{R}_{yj}+R^\phi_{yj})=
e^{-4\phi}(\tilde{R}^x_y+\tilde{\gamma}^{xj}R^\phi_{yj})$. As far
as the principal terms are concerned, we have
\begin{equation}
\tilde{\gamma}^{xj}R^\phi_{yj} =
-2\tilde{D}^x\tilde{D}_y\phi+\ldots
\end{equation}

\noindent Also as far as principal terms are concerned, the Ricci
tensor of a metric of unit determinant is
\begin{equation}
\tilde{R}_{ij} =
-\frac12\tilde{\gamma}^{kl}(\tilde{\gamma}_{ij,kl}
    -\tilde{\gamma}_{il,jk} -\tilde{\gamma}_{jk,il}) +\ldots
\end{equation}

\noindent Thus
\begin{eqnarray}
\tilde{R}^x_y &=&
-\frac12\tilde{\gamma}^{xj}\tilde{\gamma}^{kl}(\tilde{\gamma}_{yj,kl}
    -\tilde{\gamma}_{yl,jk} -\tilde{\gamma}_{jk,yl})
    +\ldots \\
    &=& -\frac12(\tilde{\gamma}^{xm}\tilde{\gamma}^{kl}
        -\tilde{\gamma}^{xl}\tilde{\gamma}^{km})\tilde{\gamma}_{ym,kl}\nonumber\\
        &&
        +\frac12\tilde{\gamma}^{xm}\tilde{\gamma}^{kl}\tilde{\gamma}_{km,yl}
        +\ldots
\end{eqnarray}

\noindent The last term can be substituted in terms of the
first-order variable $\tilde{\Gamma}^i$:
\begin{equation}
\frac12\tilde{\gamma}^{xm}\tilde{\gamma}^{kl}\tilde{\gamma}_{km,yl}
= \frac12 \tilde{\Gamma}^x,_y +\dots
\end{equation}

\noindent Additionally,  by expanding the three values of the
index $k$ in the first term one can see that $k=x$ vanishes identically, so 
that the first term contains no second $x-$derivatives of $\tilde{\gamma}_{ij}$
as it stands, and we have explicitly
\begin{eqnarray}\label{Rxy}
\tilde{R}^x_y &=&\frac12(\tilde{\gamma}^{xl}\tilde{\gamma}^{ym}
-\tilde{\gamma}^{xm}\tilde{\gamma}^{yl})\tilde{\gamma}_{ym,ly}
\nonumber\\
&& + \frac12(\tilde{\gamma}^{xl}\tilde{\gamma}^{zm}
-\tilde{\gamma}^{xm}\tilde{\gamma}^{zl})\tilde{\gamma}_{ym,lz}
\nonumber\\
&& +\frac12 \tilde{\Gamma}^x,_y +\dots
\end{eqnarray}

\noindent On the other hand, $K^x_y = \gamma^{xm}K_{my} =
\tilde{\gamma}^{xm}(\tilde{A}_{my}+\frac13\tilde{\gamma}_{ym}K)=\tilde{A}^x_y$,
so that
\begin{equation}
\dot{K}^x_y = \dot{\tilde{A}}^x_y+\ldots
\end{equation}

\noindent Finally, thus, the principal terms of $G^x_y$ are written out
in the intended form:
\begin{eqnarray}\label{Gxybssn}
G_y^x &=& -\frac{\dot{\tilde{A}}^x_y}{\alpha}
          + e^{-4\phi}\Big(\frac12(\tilde{\gamma}^{xl}\tilde{\gamma}^{ym}
-\tilde{\gamma}^{xm}\tilde{\gamma}^{yl})\tilde{\gamma}_{ym,ly}
\nonumber\\
&& + \frac12(\tilde{\gamma}^{xl}\tilde{\gamma}^{zm}
-\tilde{\gamma}^{xm}\tilde{\gamma}^{zl})\tilde{\gamma}_{ym,lz}
 +\frac12 \tilde{\Gamma}^x,_y\nonumber\\
&& -2\tilde{D}^x\tilde{D}_y\phi\Big)+\ldots
\end{eqnarray}

By a completely analogous calculation we also have
\begin{eqnarray}\label{Gxzbssn}
G_z^x &=& -\frac{\dot{\tilde{A}}^x_z}{\alpha}
          + e^{-4\phi}\Big(\frac12(\tilde{\gamma}^{xl}\tilde{\gamma}^{ym}
-\tilde{\gamma}^{xm}\tilde{\gamma}^{yl})\tilde{\gamma}_{zm,ly}
\nonumber\\
&& + \frac12(\tilde{\gamma}^{xl}\tilde{\gamma}^{zm}
-\tilde{\gamma}^{xm}\tilde{\gamma}^{zl})\tilde{\gamma}_{zm,lz}
 +\frac12 \tilde{\Gamma}^x,_z\nonumber\\
&& -2\tilde{D}^x\tilde{D}_z\phi\Big)+\ldots
\end{eqnarray}

We can treat $G^x_x$ with a similar procedure. By analogy with
(\ref{Rxy}) the reader should have no trouble to see that
\begin{eqnarray}
\tilde{R}^x_x &=&\frac12(\tilde{\gamma}^{xl}\tilde{\gamma}^{ym}
-\tilde{\gamma}^{xm}\tilde{\gamma}^{yl})\tilde{\gamma}_{xm,ly}
\nonumber\\
&& + \frac12(\tilde{\gamma}^{xl}\tilde{\gamma}^{zm}
-\tilde{\gamma}^{xm}\tilde{\gamma}^{zl})\tilde{\gamma}_{xm,lz}
\nonumber\\
&& +\frac12 \tilde{\Gamma}^x,_x +\dots
\end{eqnarray}

\noindent But since
\begin{eqnarray}
\tilde{R} &=& \tilde{\gamma}^{ij}\tilde{R}_{ij} =
\tilde{\gamma}^{ij}\tilde{\gamma}^{kl}\tilde{\gamma}_{il,kj}+\ldots\nonumber\\
&=&-\tilde{\gamma}^{kj},_{kj} +\ldots \nonumber\\
&=& \tilde{\Gamma}^k,_k +\ldots
\end{eqnarray}

\noindent then the term with an $x-$derivative of
$\tilde{\Gamma}^x$ in $\tilde{R}^x_x$ cancels out with a term in
$-\frac12\tilde{R}$, so that:
\begin{eqnarray}
\tilde{R}^x_x -\frac12 \tilde{R}
&=&\frac12(\tilde{\gamma}^{xl}\tilde{\gamma}^{ym}
-\tilde{\gamma}^{xm}\tilde{\gamma}^{yl})\tilde{\gamma}_{xm,ly}
\nonumber\\
&& + \frac12(\tilde{\gamma}^{xl}\tilde{\gamma}^{zm}
-\tilde{\gamma}^{xm}\tilde{\gamma}^{zl})\tilde{\gamma}_{xm,lz}
\nonumber\\
&& -\frac12 \tilde{\Gamma}^y,_y -\frac12 \tilde{\Gamma}^z,_z
+\dots
\end{eqnarray}

\noindent Additionally, we have $K^x_x = \tilde{A}^x_x + \frac13K$, so that
\begin{equation}
\dot{K}^x_x = \dot{\tilde{A}}^x_x+\frac13 \dot{K} + \ldots
\end{equation}

\noindent Collecting the relevant results, the principal terms of $G^x_x$ are
finaly  written out in the intended form:
\begin{eqnarray}\label{Gxxbssn}
G_x^x &=&-\frac{\dot{\tilde{A}}^x_x}{\alpha}
+\frac{2}{3\alpha}\dot{K}
%\nonumber\\ &&
+e^{-4\phi}\Big( \frac12(\tilde{\gamma}^{xl}\tilde{\gamma}^{ym}
-\tilde{\gamma}^{xm}\tilde{\gamma}^{yl})\tilde{\gamma}_{xm,ly}
\nonumber\\ &&
+ \frac12(\tilde{\gamma}^{xl}\tilde{\gamma}^{zm}
-\tilde{\gamma}^{xm}\tilde{\gamma}^{zl})\tilde{\gamma}_{xm,lz}
%\nonumber\\ &&
-\frac12 \tilde{\Gamma}^y,_y -\frac12
\tilde{\Gamma}^z,_z\nonumber\\
&& + 2\tilde{D}^y\tilde{D}_y\phi +
 2\tilde{D}^z\tilde{D}_z\phi \Big) +\ldots
 \end{eqnarray}

For the remaining boundary equation, $G^x_t$, we observe that
\begin{equation}
G^x_t = -\frac12\gamma^{xi}\big( (\ln(\det\gamma_{kl})),_{it}
    -\gamma^{kl}\dot{\gamma}_{ik,l}\big) + \ldots
\end{equation}

\noindent which, with the simple substitution of
$\gamma_{ij}=e^{4\phi}\tilde{\gamma}_{ij}$, becomes
\begin{equation}\label{Gxtbssn}
G^x_t = -\frac12e^{-4\phi}\big(8\tilde{\gamma}^{xi} \dot{\phi},_i
+\dot{\gamma}^{xl},_l\big) + \ldots
\end{equation}

\noindent which we take as the intended form without further
substitutions.

In principle, thus, we have four boundary equations 
\begin{subequations}
\begin{eqnarray}
G^x_t &=& 0,\\
G^x_x &=& 0,\\
G^x_y &=& 0,\\
G^x_z &=& 0,
\end{eqnarray}
\end{subequations}

\noindent with $G^x_y, G^x_z, G^x_x, G^x_t$ given by (\ref{Gxybssn}),
(\ref{Gxzbssn}), (\ref{Gxxbssn}) and (\ref{Gxtbssn}), as candidates for
nontrivial boundary conditions for the evolution equations (\ref{bssn}). Notice
that, unlike the ADM case, none of the four equations is manifestly identical
to any of the evolution equations (\ref{bssn}), the reason being that all the
evolution equations for the traceless extrinsic curvature density
$\tilde{A}_{ij}$ contain second derivatives of $\tilde{\gamma}_{ij}$ as part of
the Laplacian operator $\tilde{\gamma}^{kl}\partial_k\partial_l$. This makes it
a nontrivial task to figure out whether the boundary equations are identically
satisfied by the solution of the evolution equations with constrained initial
data. In the following Section, we argue that at least three
of them are necessary, on the basis of their role in ensuring the propagation
of the vanishing values of the constraints.

%-------------------------------------------------------------------

\section{Relation to constraint propagation\label{sec:4}}

The scalar and vector constraints ${\cal C}$ and ${\cal C}_i$ of
(\ref{admconst}) can be expressed as follows in terms of the BSSN variables:
\begin{subequations}\label{bssnconst}
\begin{eqnarray}
    {\cal C} &\equiv &e^{-4\phi}\left(\tilde{\Gamma}^k,_k
		-8\tilde{D}^l\tilde{D}_l\phi
		-8\tilde{D}^l\phi\tilde{D}_l\phi\right.\nonumber\\
&& \left.
		+\frac14\tilde{\gamma}^{lm}\tilde{\gamma}^{jk},_l
		(\tilde{\gamma}_{jk,m}-2\tilde{\gamma}_{jm,k})\right)\nonumber\\
&&
		- \tilde{A}_{ij}\tilde{A}^{ij} -\frac23 K^2 ,   \label{cbssn}\\
    {\cal C}_i&\equiv & \tilde{\gamma}^{jk}\tilde{A}_{ki,j}
	-\tilde{\Gamma}^k\tilde{A}_{ik} 
	-\frac12\tilde{\gamma}_{jk,i}\tilde{A}^{jk}
	-\frac23 K,_i \nonumber\\
&& +6\tilde{A}^j_i\phi,_j
\end{eqnarray}

\noindent In addition to the scalar and vector constraints ${\cal C}$ and ${\cal C}_i$,
an additional constraint,
\begin{equation}
{\cal G}^i \equiv \tilde{\Gamma}^i+\tilde{\gamma}^{ij},_j
\end{equation}
\end{subequations}

\noindent must be imposed on the initial values, i.e., 
\begin{equation}
	 {\cal G}^i =0
\end{equation}

\noindent in order for the BSSN equations (\ref{bssn}) to yield a solution of
the ADM equations (\ref{adm}). Whether the constraints ${\cal C}, {\cal C}_i$
and ${\cal G}^i$ are satisfied by the solution of the evolution equations
(\ref{bssn}) is determined by the equations for $\dot{\cal C}, \dot{\cal C}_i$
and $\dot{\cal G}^i$ that are implied by (\ref{bssn}). The easiest propagation
equation to obtain is that for ${\cal G}^i$, because ${\cal G}^i$ is a trivial
constraint, whose propagation should be exactly zero by construction unless
some constraint had been used in establishing the evolution equations for
$\tilde{\Gamma}^i$ or for $\tilde{\gamma}_{ij}$. Since the vector constraint is
necessary to write Eq.~(\ref{dotGamma}) in such a form, the evolution of ${\cal
G}^i$ naturally depends on the vector constraint, as follows:
\begin{equation} 
\dot{\cal G}^i = 2\alpha\tilde{\gamma}^{ij}{\cal C}_j
\end{equation}

\noindent This equation has no principal terms for the purposes of the
structure of characteristics, so we can write it as
\begin{equation} \label{gdotbssn}
\dot{\cal G}^i = \ldots
\end{equation}

\noindent where $\ldots$ represents undifferentiated terms. 
Additionally, since the time-derivatives of the variables that appear in the
expression (\ref{cbssn}) for ${\cal C}$ will not contribute any terms that would
have derivatives of the constraints themselves, we have
\begin{equation}\label{cdotbssn}
\dot{\cal C} = \ldots
\end{equation} 

\noindent as well. Finally, the only term that contributes derivatives of
constraints to $\dot{\cal C}_i$ is
$\tilde{\gamma}^{jk}\dot{\tilde{A}}_{ki,j}$, yielding
\begin{equation}\label{cidotbssn}
\dot{\cal C}_i = \frac{\alpha e^{-4\phi}}{2}\tilde{\gamma}_{ik}
		\tilde{D}^l\tilde{D}_l{\cal G}^k 
		+\frac{\alpha}{6} {\cal C},_i +\ldots
\end{equation}

\noindent As it stands, the system of evolution equations (\ref{gdotbssn}),
(\ref{cdotbssn}) and (\ref{cidotbssn}) is not manifestly well posed, not even
manifestly hyperbolic, because of the presence of second space derivatives in
the right-hand side as a direct consequence of the addition of the three
first-order variables $\tilde{\Gamma}^i$. In order to facilitate the study of
the properties of propagation of the constraints, however, we may define a set
of six first-order variables for the constraints that allow us to reduce the
differential order of the right-hand side: 
\begin{equation}
{\cal I}_{ik} \equiv \tilde{\gamma}_{il} {\cal G}^l,_k
\end{equation}

\noindent These are not to be thought of as additional constraints to be imposed
on the initial data of the BSSN evolution equations. Rather, they will be
automatically satisfied by any set of initial data that satisfies ${\cal
G}^i=0$. But they are useful because, by adding them as new variables,  
the propagation of the constraints takes the form:
\begin{subequations}\label{propbssn}
\begin{eqnarray}
\dot{\cal C} &=& \ldots\\
\dot{\cal C}_i &=& \frac{\alpha}{6} {\cal C},_i
		+\frac{\alpha e^{-4\phi}}{2}\tilde{\gamma}^{kl}
		{\cal I}_{ik,l} +\ldots\\
\dot{\cal I}_{ik} &=& 2\alpha {\cal C}_{i,k} + \ldots\\
\dot{\cal G}^i &=& \ldots
\end{eqnarray}
\end{subequations}

\noindent This first-order system of 13 equations is strongly hyperbolic in the
sense that it has a complete set of null eigenvectors. This means that for each
arbitrary covector $\xi_i$ of unit length $\xi_i\xi_j\gamma^{ij}=1$ at any given
point, there are 13 characteristic fields, which travel with real speeds. One
can easily see that the characteristic speeds of this system are $0$, and
$\pm \alpha$, with multiplicity 7, 3 and 3 respectively.  Thus seven
characteristic constraints are ``static'' in the sense that they propagate
along the direction of evolution $\partial/\partial t$.  Of the six non-static
characteristic constraints, three are outgoing at the boundary:
\begin{equation}\label{Z+}
	{}^+{\cal Z}_i \equiv {\cal C}_i
	+\frac12 \xi^k{\cal  I}_{ik} + \frac16 \xi_i {\cal C} 
\end{equation}

\noindent and three are incoming:
\begin{equation}\label{Z-}
	{}^-{\cal Z}_i \equiv {\cal C}_i
	-\frac12 \xi^k{\cal  I}_{ik} - \frac16 \xi_i {\cal C} 
\end{equation}

\noindent This indicates that, even if the 13 constraints are set to zero on
the initial slice, they would not remain zero in the future of the initial
slice unless the incoming constraints are set to zero, either directly or
indirectly as functions of the outgoing constraints. One way to accomplish this
would be by means of boundary conditions of the form ${}^-{\cal Z}_i = L_{ij}
{}^+{\cal Z}_j$ with constant coefficients $L_{ij}$, which is a necessary
condition in order to preserve the well-posedness of the propagation of the
constraints~\cite{kreissbook}. 

Our next goal is to show that the boundary equations $G^x_a=0$ with $G^x_a$
given by (\ref{Gxybssn}),(\ref{Gxzbssn}), (\ref{Gxxbssn}) and (\ref{Gxtbssn})
are equivalent, in a sense, to such boundary conditions. This
would indicate, in effect, that the boundary equations $G^x_a=0$ are
instrumental in preserving the constraints. 

More precisely, we wish to show that the boundary equations $G^x_a=0$ are
related to the non-static constraints through linear combinations with the
evolution equations in the following manner:
\begin{subequations}\label{38}
\begin{eqnarray}
G^x_t &=& -\alpha {\cal C}^x -\frac12\gamma^{xl}\left(
	8\tilde{\cal E}^\phi,_l-\tilde{\gamma}^{kj}
	\tilde{\cal E}^\gamma_{lk,j}\right)\label{38a}\\
G^x_x &=& -\left(\frac16 {\cal C}+\frac12 \xi^m{\cal I}_{xm}\right)
	-\frac{1}{\alpha}\tilde{\gamma}^{xj}\tilde{\cal E}^A_{xj} 
	+\frac{2}{3\alpha} \tilde{\cal E}^K\hspace{1cm}\label{38b}\\
G^x_y &=& -\frac12 \xi^m{\cal I}_{ym}
	-\frac{1}{\alpha}\gamma^{xj}\tilde{\cal E}^A_{yj} \label{38c}\\
G^x_z &=& -\frac12 \xi^m{\cal I}_{zm}
	-\frac{1}{\alpha}\gamma^{xj}\tilde{\cal E}^A_{zj} \label{38d}
\end{eqnarray}
\end{subequations}

\noindent where $\tilde{\cal E}^\phi, \tilde{\cal E}^\gamma_{lk}, \tilde{\cal
E}^A_{ij}, \tilde{\cal E}^K $ represent the evolution equations
(\ref{bssndotphi}), (\ref{bssndotg}), (\ref{bssndotA}) and
(\ref{bssndotK}) where all the terms in the right-hand side are moved to
the left. We can think of the relationship between the boundary equations and
the constraints through the evolution equations as an equivalence: 
\begin{subequations}
\begin{eqnarray}
G^x_t &\sim& -\alpha {\cal C}^x \\
G^x_x &\sim& -\left(\frac16 {\cal C}+\frac12 \xi^m{\cal I}_{xm}\right)
	\\
G^x_y &\sim& -\frac12 \xi^m{\cal I}_{ym}
	\\
G^x_z &\sim& -\frac12 \xi^m{\cal I}_{zm}
\end{eqnarray}
\end{subequations}

\noindent modulo the evolution. 

In order to prove this, we start with the boundary equation $G^x_x=0$. We have,
explicitly from (\ref{bssndotA}),
\begin{eqnarray}
\tilde{\gamma}^{xl}\tilde{\cal E}^A_{xl}
&=& \dot{\tilde{A}}^x_x - \alpha e^{-4\phi}\Big(
	-\frac12\tilde{\gamma}^{kl} \tilde{\gamma}^{xj}\tilde{\gamma}_{xj,kl}
	+\frac12 \tilde{\Gamma}^x,_x \nonumber\\
&&
	+\frac12\tilde{\gamma}^{xj}\tilde{\gamma}_{kx}\tilde{\Gamma}^k,_j
	-\frac13 \tilde{\Gamma}^k,_k
	-2\tilde{D}^x\tilde{D}_x\phi
	+\frac23 \tilde{D}^k\tilde{D}_k\phi\Big)\nonumber\\
&& +\ldots
\end{eqnarray}

\noindent and also
\begin{equation}
\tilde{\gamma}^{xl}\tilde{\cal E}^K_{xl} = \dot{K}+ \ldots
\end{equation}

Therefore, adding the evolution equations (with appropriate factors) to
$G^x_x$ as given by Eq.~(\ref{Gxxbssn}), we have immediately
\begin{eqnarray}\label{preGxxevol}
\lefteqn{G^x_x +\frac{1}{\alpha}\tilde{\gamma}^{xj}\tilde{\cal E}^A_{xj} 
	-\frac{2}{3\alpha} \tilde{\cal E}^K} &&\\  
&=& \frac{e^{-4\phi}}{2}\Big( 
	\tilde{\gamma}^{xj}\tilde{\gamma}^{kl}\tilde{\gamma}_{xj,kl}
	+(\tilde{\gamma}^{xl}\tilde{\gamma}^{ym}
-\tilde{\gamma}^{xm}\tilde{\gamma}^{yl})\tilde{\gamma}_{xm,ly}
\nonumber\\ &&
+ (\tilde{\gamma}^{xl}\tilde{\gamma}^{zm}
-\tilde{\gamma}^{xm}\tilde{\gamma}^{zl})\tilde{\gamma}_{xm,lz}
-\frac13 \tilde{\Gamma}^k,_k 
+ \frac83\tilde{D}^k\tilde{D}_k\phi \nonumber\\ &&
-\tilde{\gamma}^{xj}\tilde{\gamma}_{kx}\tilde{\Gamma}^k,_j \Big)+\ldots
\end{eqnarray} 

\noindent The three indicated terms in second derivatives of 
$\tilde{\gamma}_{ij}$ manifestly combine to yield
\begin{eqnarray}
(\tilde{\gamma}^{xl}\tilde{\gamma}^{ym}
-\tilde{\gamma}^{xm}\tilde{\gamma}^{yl})\tilde{\gamma}_{xm,ly}&&\nonumber\\
+(\tilde{\gamma}^{xl}\tilde{\gamma}^{zm}
-\tilde{\gamma}^{xm}\tilde{\gamma}^{zl})\tilde{\gamma}_{xm,lz}&&\nonumber\\
+\tilde{\gamma}^{xj}\tilde{\gamma}^{kl}\tilde{\gamma}_{xj,kl}
&=& \tilde{\gamma}^{xm}\tilde{\gamma}^{kl}\tilde{\gamma}_{xl,km}.
\end{eqnarray}

\noindent Thus (\ref{preGxxevol}) reads, equivalently,
\begin{eqnarray}\label{nextGxxevol}
G^x_x +\frac{1}{\alpha}\tilde{\gamma}^{xj}\tilde{\cal E}^A_{xj} 
	-\frac{2}{3\alpha} \tilde{\cal E}^K 
% &&\\  
&=& \Big(
\tilde{\gamma}^{xm}\big(
 \tilde{\gamma}^{kl}\tilde{\gamma}_{xl,km}
-\tilde{\gamma}_{kx}\tilde{\Gamma}^k,_m\big)\nonumber\\
&&
-\frac13\big(\tilde{\Gamma}^k,_k-8\tilde{D}^l\tilde{D}l\phi\big)\Big)
\frac{e^{-4\phi}}{2} \nonumber\\
&&+\ldots	
\end{eqnarray}

\noindent Both indicated terms are directly expressible in terms of constraints
now. The second term in parenthesis (with the exponential factor included) is
manifestly equal to the Hamiltonian constraint ${\cal C}$, whereas the first
parenthesis immediately leads to one of the new first-order constraints:
\begin{eqnarray}
\tilde{\gamma}^{kl}\tilde{\gamma}_{xl,km}
-\tilde{\gamma}_{kx}\tilde{\Gamma}^k,_m &=& 
-\tilde{\gamma}_{kx}\big(\tilde{\Gamma}^k + \tilde{\gamma}^{kj},_j\big),_m
+\ldots\nonumber\\
 &=& -\tilde{\gamma}_{kx}{\cal G}^k,_m +\ldots\nonumber\nonumber\\
 &=& -{\cal I}_{xm} +\dots
\end{eqnarray} 

\noindent Thus Eq.~(\ref{38b}) is verified. 

We now show how to prove (\ref{38c}). From (\ref{bssndotA}) we have
\begin{eqnarray}
\frac{1}{\alpha}\gamma^{xj}\tilde{\cal E}^A_{yj} &=& \frac{e^{-4\phi}}{2}\Big(
	\tilde{\gamma}^{xj}\tilde{\gamma}^{lm}\tilde{\gamma}_{yj,lm}
	-\tilde{\gamma}^{xj}\tilde{\gamma}_{yk}\tilde{\Gamma}^k,_j
	-\tilde{\Gamma}^x,_y \nonumber\\
&&
	+4\tilde{D}^x\tilde{D}_y\phi\Big)+ \dot{\tilde A}^x_y +\dots
\end{eqnarray}

\noindent Adding this to $G^x_y$ as given by (\ref{Gxybssn}) we manifestly
obtain
\begin{eqnarray}
G^x_y + \frac{1}{\alpha}\gamma^{xj}\tilde{\cal E}^A_{yj}
&=& \Big(
(\tilde{\gamma}^{xl}\tilde{\gamma}^{ym}
-\tilde{\gamma}^{xm}\tilde{\gamma}^{yl})\tilde{\gamma}_{ym,ly}\nonumber\\
&&
	+(\tilde{\gamma}^{xl}\tilde{\gamma}^{zm}
-\tilde{\gamma}^{xm}\tilde{\gamma}^{zl})\tilde{\gamma}_{ym,lz}\nonumber\\
&&	+\tilde{\gamma}^{xj}\tilde{\gamma}^{lm}\tilde{\gamma}_{yj,lm}
-\tilde{\gamma}^{xj}\tilde{\gamma}_{yk}\tilde{\Gamma}^k,_j
	\Big)\frac{e^{-4\phi}}{2}\nonumber\\
&&+\dots 
\end{eqnarray} 

\noindent By expanding the index $l=x,y,z$ in the third line of the preceeding
equation some cancellations take place, leading directly to
\begin{equation}
G^x_y + \frac{1}{\alpha}\gamma^{xj}\tilde{\cal E}^A_{yj}
= \frac{e^{-4\phi}}{2}\big(
\tilde{\gamma}^{xl}\tilde{\gamma}^{km}\tilde{\gamma}_{ym,kl}
-\tilde{\gamma}^{xj}\tilde{\gamma}_{yk}\tilde{\Gamma}^k,_j\big)
+\dots 
\end{equation} 

\noindent The terms in parenthesis in the right-hand side are now expressible in
terms of the new first-order constraint:
\begin{eqnarray}
\tilde{\gamma}^{xl}\tilde{\gamma}^{km}\tilde{\gamma}_{ym,kl}
-\tilde{\gamma}^{xj}\tilde{\gamma}_{yk}\tilde{\Gamma}^k,_j
&=& -\tilde{\gamma}^{xj}\tilde{\gamma}_{yk}\big(\tilde{\gamma}^{kl},_l
  +\tilde{\Gamma}^k\big),_j +\ldots \nonumber\\
&=& -\tilde{\gamma}^{xj}{\cal I}_{yj} +\ldots
\end{eqnarray}

\noindent Thus Eq.~(\ref{38c}) is verified.  A completely analogous procedure
(with $z$ in the place of the subscript $y$) leads to the verification of
Eq.~(\ref{38d}). 

It only remains to verify Eq.~(\ref{38a}). From (\ref{bssndotphi}) and
(\ref{bssndotg}) we have trivially
\begin{equation}
\tilde{\cal E}^\phi = \dot{\phi}+\ldots
\end{equation}

\noindent and
\begin{equation}
\tilde{\cal E}^\gamma_{lk} = \dot{\tilde{\gamma}}_{lk}+\ldots
\end{equation}

\noindent Adding these (in the manner specified by (\ref{38a})) to $G^x_t$ as
given by (\ref{Gxtbssn}) we have immediately
\begin{eqnarray}
\lefteqn{G^x_t +\frac12\gamma^{xl}\left(
	8\tilde{\cal E}^\phi,_l-\tilde{\gamma}^{kj}
	\tilde{\cal E}^\gamma_{lk,j}\right)}&&\nonumber\\
 &=& e^{-4\phi}\alpha\left(\frac23\tilde{\gamma}^{xl}K,_l
	-\tilde{\gamma}^{xl}\tilde{\gamma}^{kj}\tilde{A}_{kl,j}\right)
	+\dots
\end{eqnarray}

\noindent The right-hand side of the preceeding equation is manifestly 
expressible in terms of the vector constraint ${\cal C}_l$, with the
consequence that Eq.~(\ref{38a}) is verified. By the set of equations
(\ref{38}), thus, imposing the equations $G^x_a=0$ on the boundary is
equivalent, in a sense, to imposing the indicated combinations of constraints
on the boundary.  We now show that such combinations of constraints are
equivalent to the incoming characteristic constraints. 

By Eqs.~(\ref{38}), (\ref{Z+}) and (\ref{Z-}), using
$\xi_i=(1,0,0)/\sqrt{\gamma^{xx}}$, we have
\begin{subequations}
\begin{eqnarray}
G^x_t &\sim& -\frac{\alpha}{2} \gamma^{xi}\left({}^-{\cal Z}_i +{}^+{\cal Z}_i\right)\\
G^x_x &\sim& \frac12\left({}^-{\cal Z}_x -{}^+{\cal Z}_x\right)
	\\
G^x_y &\sim& \frac12\left({}^-{\cal Z}_y -{}^+{\cal Z}_y\right)
	\\
G^x_z &\sim& \frac12\left({}^-{\cal Z}_z -{}^+{\cal Z}_z\right)
\end{eqnarray}
\end{subequations}

\noindent The four components $G^x_a$ are equivalent (modulo the evolution) to
linear combinations of the incoming and outgoing constraints with constant
coefficients. Thus imposing $G^x_a=0$ is equivalent (modulo evolution) to
setting well-posed boundary conditions on the system of propagation
of the constraints (${}^-{\cal Z}_i = L_i^j{}^+{\cal Z}_j$).  The propagation
of the constraints requires only three boundary conditions.  Accordingly,
three linearly independent combinations of the four equations $G^x_a=0$ are
necessary in order to ensure that the constraints will remain satisfied by the
solution of the evolution equations (\ref{bssn}) with constrained initial
data. A fourth lindearly independent one should be satisfied identically.  

This means that by imposing three (linearly independent combinations)
of the four equations $G^x_a=0$ on the boundary values of the evolution
equations (\ref{bssn}) the propagation of the constraints is enforced.     

At this level there is no preferred set of three linear combinations out of
the four equations $G^x_a=0$. This freedom is an advantage, because it may
easily accomodate requirements that may stem from expectations of numerical
stability. We do not concern ourselves with such considerations in this work.

The question arises as to whether the fourth equation should be used on the
boundary values. For the sake of argument, we can take the fourth equation to
be $G^x_t+\alpha \gamma^{xi}G^x_i=0$, since this equation is equivalent to a
linear combination of purely outgoing constraints (${}^+{\cal Z}_i\gamma^{xi}
=0$).  If one looks at this question from the point of view of the
constraint-propagation problem (\ref{propbssn}), clearly imposing the fourth
equation would be redundant, given that the outgoing constraints are made to
vanish at the boundary simply by picking vanishing initial values. Yet the
fourth equation is not \textit{inconsistent} with the choice of vanishing
initial values for the outgoing constraints.  

From the point of view of the evolution equations (\ref{bssn}), however, it is
not at all clear that the fourth equation should not be imposed. This is
because, as it stands, the system of evolution equations (\ref{bssn}) is not
strongly hyperbolic in the standard sense (see, however, \cite{NOR}).
Therefore, there is no clear classification of the 15 fundamental variables
into static, incoming or outgoing fields. As a consequence, to our knowledge,
the actual number of boundary conditions required by the evolution equations
(\ref{bssn}), being not less than three, remains unknown. 

In the next section we reduce the BSSN equations fully to first-order form. The
resulting first-order system is strongly hyperbolic in the standard sense and
the boundary-value-problem can be solved in the standard manner.    

%-------------------------------------------------------------------
\section{Einstein boundary conditions for a well-posed first-order reduction of
the BSSN equations}\label{sec:5}

In order for Eqs.~(\ref{bssn}) to reduce to first-order form, the spatial
derivatives of $\phi$ and those of $\tilde{\gamma}_{ij}$ not already accounted
for by $\tilde{\Gamma}^i$ need to be added as new variables (a total of 15 new
variables).  They are
\begin{subequations}
\begin{eqnarray}
Q_i &\equiv& \phi,_i \label{Qdef}\\
V_{ijk} &\equiv& \tilde{\gamma}_{ij,k} 
-\frac35\tilde{\gamma}_{k(i}\tilde{\gamma}_{j)l,m}\tilde{\gamma}^{lm}
+\frac15\tilde{\gamma}_{ij}\tilde{\gamma}_{kl,m}\tilde{\gamma}^{lm} \label{Vdef}
\end{eqnarray}
\end{subequations}
	
\noindent These new first-order variables have been used before in order to
find a symmetrizable hyperbolic first-order reduction of the BSSN equations by
linear combination with the constraints~\cite{simo99}. The difference here is
that we do not involve combinations with the constraints.  By using
$\tilde{\gamma}^{ij}$ to raise the first two
indices in (\ref{Vdef}) we have
\begin{equation}
V^{ij}{}_k = -\tilde{\gamma}^{ij},_k 
+\frac35\delta^{(i}_k\tilde{\gamma}^{j)m},_m
-\frac15\tilde{\gamma}^{ij}\tilde{\gamma}^{lm},_m\tilde{\gamma}_{kl}
\end{equation} 

\noindent One has $V^{ij}{}_j=0$ automatically, so the set of variables
$V^{ij}{}_k$ does not include the three variables $\tilde{\Gamma}^i$.
Additionally,  $V^{ij}{}_k$ is traceless in the first two indices, also by
construction ($V^{ij}{}_k\tilde{\gamma}_{ij}=0$).  Thus  $V^{ij}{}_k$ are 12
variables in all. Using $V^{ij}{}_k$ and $\tilde{\Gamma}^i$ one is able to
invert in order to obtain all the derivatives of the inverse metric in terms of
the new variables:
\begin{equation}
\tilde{\gamma}^{ij},_k = -V^{ij}{}_k 
-\frac35\delta^{(i}_k\tilde{\Gamma}^{j)}
+\frac15\tilde{\gamma}^{ij}\tilde{\Gamma}^l\tilde{\gamma}_{kl}
\end{equation}

In addition to reducing the BSSN equations to first-order form, we also require
the lapse function to be densitized, namely: 
\begin{equation}
\alpha = \sigma\sqrt{\det(\gamma_{ij})} = \sigma e^{6\phi}
\end{equation}

\noindent where $\sigma$ is an arbitrarily prescribed function of spacetime. The
evolution equations (\ref{bssn}), augmented by trivial evolution equations
for $Q_i$ and $V^{ij}{}_k$, read now:
\begin{subequations}\label{bssn1}
\begin{eqnarray}
\dot{\tilde{\gamma}}_{ij} &=& -2\alpha \tilde{A}_{ij} \label{bssn1dotg}\\
\dot{\phi} &=& -\frac{\alpha}{6} K \label{bssn1dotphi}\\
\dot{\tilde{A}}_{ij} &=&-\Big(
\frac12 \tilde{\gamma}_{ir}\tilde{\gamma}_{js}\tilde{\gamma}^{km}V^{rs}{}_{k,m}
-\frac{7}{10}\tilde{\gamma}_{k(i}\tilde{\Gamma}^k,_{j)}
+\frac{7}{30}\tilde{\gamma}_{ij}\tilde{\Gamma}^k,_k \nonumber\\
&&
+8 Q_{(i,j)}
-\frac83\tilde{\gamma}_{ij}\tilde{\gamma}^{kl}Q_{k,l} \Big)
\alpha e^{-4\phi}+\ldots \label{bssn1dotA}
\\
%\end{eqnarray}
%\begin{eqnarray}
\dot{K} &=& -6\alpha\tilde{\gamma}^{kl}Q_{k,l}+\ldots\label{bssn1dotK}\\
\dot{\tilde{\Gamma}}^i &=& -\frac43\alpha\tilde{\gamma}^{ij}K,_j
+\ldots\label{dot1Gamma}\\
\dot{Q}_i &=& -\frac{\alpha}{6}K,_i +\ldots \label{dotQ}\\
\dot{V}^{ij}{}_k &=& -2\alpha \Big(
\tilde{A}^{ij},_k
-\frac35 \delta^{(i}\tilde{A}^{j)s},_s
+\frac15 \tilde{\gamma}^{ij}\tilde{\gamma}_{kl}\tilde{A}^{ls},_s\Big)\nonumber\\
&&
+\ldots \label{dotV}
\end{eqnarray}
\end{subequations}

\noindent The new evolution equations (\ref{dotQ}) and (\ref{dotV}) are
obtained simply by taking a time derivative of the definitions (\ref{Qdef}) and
(\ref{Vdef}) and substituting the evolution equations (\ref{bssndotphi}) and
(\ref{bssndotg}) in the resulting right-hand sides. No linear combinations with
the constraints are used. 

In order to see that this initial value problem is strongly hyperbolic in the
santadard sense~\cite{kreissbook}, the eigenvalue problem of the principal
symbol in an arbitrary direction must be solved. This is equivalent to assuming
that all the fields have a plane wave dependence of the form $\exp
i(\xi_kx^k-st)$ with real $s$ and arbitrary $\xi_i$ so that
$\xi_i\xi_j\gamma^{ij}=1$.  The eigenvalue problem of the principal symbol has
thus the form:
\begin{subequations}\label{bssn1eigen}
\begin{eqnarray}
s\tilde{\gamma}_{ij} &=& 0\\
s\phi &=&0\\
s\tilde{A}_{ij} &=&-\alpha e^{-4\phi}\Big(
\frac12 \tilde{\gamma}_{ir}\tilde{\gamma}_{js}\tilde{\gamma}^{km}V^{rs}{}_k\xi_m
-\frac{7}{10}\tilde{\gamma}_{k(i}\xi_{j)}\tilde{\Gamma}^k\nonumber\\
&&
+\frac{7}{30}\tilde{\gamma}_{ij}\tilde{\Gamma}^k\xi_k 
+8 Q_{(i}\xi_{j)}
-\frac83\tilde{\gamma}_{ij}\tilde{\gamma}^{kl}Q_k\xi_l \Big)
\\
%\end{eqnarray}
%\begin{eqnarray}
sK &=& -6\alpha\tilde{\gamma}^{kl}Q_k\xi_l\\
s\tilde{\Gamma}^i &=& -\frac43\alpha\tilde{\gamma}^{ij}K\xi_j
\\
sQ_i &=& -\frac{\alpha}{6}K\xi_i \\
sV^{ij}{}_k &=& -2\alpha \Big(
\tilde{A}^{ij}\xi_k
-\frac35 \delta^{(i}\tilde{A}^{j)s}\xi_s
+\frac15 \tilde{\gamma}^{ij}\tilde{\gamma}_{kl}\tilde{A}^{ls}\xi_s\Big)\nonumber\\
&&
\end{eqnarray}
\end{subequations}

The reader can verify that this problem has the following eigenvalues:
\begin{itemize}
\item $s=0$ with 18 eigenvectors labeled by free values of $\tilde{\gamma}_{ij},
\phi, \tilde{\Gamma}^i,$ the two components of $Q_i$ except $\gamma^{kl}Q_k \xi_l$ and the seven 
components of $V^{ij}{}_k$ \textit{other than} $\gamma^{kl}V^{ij}{}_k\xi_l$. 
\item $s=\pm\alpha$ with eigenvectors given by
\begin{eqnarray*}
\tilde{\gamma}_{ij}&=&0\\
\phi&=&0\\
\tilde{\Gamma}^i&=& -\frac{4\alpha}{3s}e^{4\phi}\xi^iK\\
Q_i&=&-\frac{\alpha}{6s}\xi_iK\\
\tilde{A}^{ij}\xi_i &=&\frac{2s}{3}e^{8\phi}\xi^iK \\
V^{ij}{}_k&=&\frac{4\alpha}{5}\Big(\delta^{(i}\xi^{j)}
-\frac13 \gamma^{ij}\xi_k\Big)e^{4\phi}K
-\frac{2\alpha}{s}\tilde{A}^{ij}\xi_k
\end{eqnarray*} 

\noindent with free values of $K$ and of the two components of $\tilde{A}^{ij}$
\textit{other than\/} $\tilde{A}^{ij}\xi_j$. These are six eigenvectors in all:
three for $s=\alpha$ and three for $s=-\alpha$. 
\item $s=\pm\alpha\sqrt{3/5}$ with
eigenvectors given by 
\begin{eqnarray*}
\tilde{\gamma}_{ij}&=&0\\
\phi&=&0\\
\tilde{\Gamma}^i&=& 0\\
Q_i&=&0\\
K &=& 0\\
\tilde{A}^{ij}&=&\frac32 \Big(\xi^i\xi^j-\frac13\gamma^{ij}\Big)
\tilde{A}^{kl}\xi_k\xi_l\\
V^{ij}{}_k&=&\frac{\alpha}{s}\Big((3\xi^i\xi^j-\gamma^{ij})\xi_k
+\frac65\delta_k^{(i}\xi^{j)}\Big)\tilde{A}^{kl}\xi_k\xi_l
\end{eqnarray*} 

\noindent with free values of the component $\tilde{A}^{kl}\xi_k\xi_l$.  These
are two eigenvectors in all: one for $s=\alpha\sqrt{3/5}$ and one for 
$s=-\alpha\sqrt{3/5}$. 

\item $s=\pm\alpha\sqrt{7/10}$ with eigenvectors given by
\begin{eqnarray*}
\tilde{\gamma}_{ij}&=&0\\
\phi&=&0\\
\tilde{\Gamma}^i&=& 0\\
Q_i&=&0\\
K &=& 0\\
\tilde{A}^{ij}&=&2\xi^{(i}\tilde{A}^{j)m}\xi_m\\
V^{ij}{}_k&=&-\frac{\alpha}{s}\Big(
4\xi^{(i}\tilde{A}^{j)m}\xi_m\xi_k
-\frac{12}{5}\delta_k^{(i}\tilde{A}^{j)m}\xi_m\nonumber\\
&&
+\frac25\tilde{\gamma}^{ij}\tilde{\gamma}_{kl}\tilde{A}^{lm}\xi_m\Big)
\end{eqnarray*}  

\noindent with free values of the two components $\tilde{A}^{ij}\xi_j$ other
than $\tilde{A}^{kl}\xi_k\xi_l$ (which vanishes). These are four eigenvectors
in all: two for  $s=\alpha\sqrt{7/10}$ and two for  $s=-\alpha\sqrt{7/10}$.
\end{itemize}

\noindent In all of the above, as in what follows, we are using the notation 
$\xi^i \equiv \gamma^{ij}\xi_j$. 

All eigenvalues are thus real and there are $18+6+2+4=30$ linearly independent
eigenvectors. The initial value problem of Eqs.~(\ref{bssn1}) is, thus,
strongly hyperbolic in the standard sense, and has six pairs of ``non-static''
characteristic fields traveling with non-zero speeds.  Not all characteristic
speeds are equal to the speed of light, though: there are only three pairs of
incoming-outgoing characteristic fields traveling at the speed of light. The
other three pairs of ``non-static'' characteristic fields travel at subluminal
speeds. (This does not necessarily contradict \cite{lsubssn}, since in that
reference the authors use a different first-order reduction of the BSSN
equations.)  In the following we write down explicitly the characteristic fields
with non-vanishing characteristic speeds, with the aim of relating some of them
to the boundary equations of Section~\ref{sec:3}.

The three pairs of characteristic fields that travel at light speed
$s=\pm\alpha$ are
\begin{eqnarray}
{}^\pm U^K &\equiv& K \pm 6 \xi^kQ_k\\
{}^\pm U^{ij} &\equiv& 
  \tilde{A}^{ij}
-2\xi^{(i}\tilde{A}^{j)k}\xi_k
+\frac12(\xi^i\xi^j+\gamma^{ij})\tilde{A}^{kl}\xi_k\xi_l\nonumber\\
&& 
\pm\Big(
\!\!-\frac12V^{ij}{}_k\xi^k
+\xi^{(i}V^{j)l}{}_k\xi_l\xi^k
\nonumber\\&&
-\frac14(\xi^i\xi^j+\gamma^{ij})V^{ml}{}_k\xi_m\xi_l\xi^k\Big)
\end{eqnarray}

\noindent Notice that ${}^\pm U^{ij}\tilde{\gamma}_{ij}={}^\pm U^{ij}\xi_j=0$.
As observed previously, they are labeled by free values of $K$ and of the two
components of $\tilde{A}^{ij}$ other than the projections along the
characteristic direction $\xi_i$.

The pair of characteristic fields with $s=\pm\alpha\sqrt{3/5}$ is 
\begin{eqnarray}
{}^\pm U &\equiv& 
 \tilde{A}^{ij}\xi_i\xi_j -\frac23e^{4\phi}K
\pm \sqrt{\frac53}\Big(
-\frac12 V^{ml}{}_k\xi_m\xi_l\xi^k\nonumber\\
&&
+\frac{7}{15} \tilde{\Gamma}^i\xi_i
-\frac43 e^{4\phi}\xi^kQ_k\Big)
\end{eqnarray}

\noindent As observed before, it is labeled by the projection
$\tilde{A}^{ij}\xi_i\xi_k$. 

The two pairs of characteristic fields with $s=\pm \alpha \sqrt{7/10}$ are 
\begin{eqnarray}
{}^\pm U^i &\equiv& 
\tilde{A}^{ij}\xi_j -\xi^i\tilde{A}^{kl}\xi_k\xi_l
\pm\frac12\sqrt{\frac{7}{10}}\Big(\tilde{\Gamma}^i-\xi^i\tilde{\Gamma}^k\xi_k\nonumber\\
&&
-V^{ij}{}_k\xi_j\xi^k
		      +\xi^i V^{ml}{}_k\xi_m\xi_l\xi^k\Big)\nonumber\\
&&
\pm 4
\sqrt{\frac{10}{7}}(-\tilde{\gamma}^{ik}Q_k+e^{4\phi}\xi^i\xi^kQ_k)
\end{eqnarray}

\noindent Notice that $U^i\xi_i=0$.  These are labeled by the two projections
$\tilde{A}^{ij}\xi_j$ other than $\tilde{A}^{kl}\xi_k\xi_l$, as anticipated
previously. 

In order to see which of these fields are prescribed by the boundary
equations (\ref{Gxybssn}), (\ref{Gxzbssn}), (\ref{Gxxbssn}) and (\ref{Gxtbssn}),
we restrict attention to $\xi_i=(1,0,0)/\sqrt{\gamma^{xx}}$. Lowering the index $i$
with $\tilde{\gamma}_{ij}$ we have
\begin{subequations}
\begin{eqnarray}
{}^\pm U_y &=& 
\frac{\tilde{A}^x_y}{\sqrt{\gamma^{xx}}}
\pm\frac12\sqrt{\frac{7}{10}}\tilde{\gamma}_{yk}
\Big(\tilde{\Gamma}^k
-V^{xk}{}_l\frac{\gamma^{xk}}{\gamma^{xx}}\Big)\mp 4
\sqrt{\frac{10}{7}}Q_y\nonumber\\
&&\\
{}^\pm U_z &=& 
\frac{\tilde{A}^x_z}{\sqrt{\gamma^{xx}}}
\pm\frac12\sqrt{\frac{7}{10}}\tilde{\gamma}_{zk}
\Big(\tilde{\Gamma}^k
-V^{xk}{}_l\frac{\gamma^{xk}}{\gamma^{xx}}\Big)\mp 4
\sqrt{\frac{10}{7}}Q_z\nonumber\\
\end{eqnarray}
\end{subequations}

\noindent Inverting for $\tilde{A}^x_y$ and $\tilde{A}^x_z$ we have
\begin{subequations}\label{AstoUs}
\begin{eqnarray}
\tilde{A}^x_y&=& \frac{\sqrt{\gamma^{xx}}}{2}({}^- U_y + {}^+ U_y)\\
\tilde{A}^x_z&=& \frac{\sqrt{\gamma^{xx}}}{2}({}^- U_z + {}^+ U_z)
\end{eqnarray}
\end{subequations}

\noindent Clearly, the boundary equations (\ref{Gxybssn}) and (\ref{Gxzbssn})
prescribe the time derivatives of the incoming fields ${}^- U_y$ and ${}^-
U_z$, respectively, in terms of their outgoing counterparts. 

Next, with $\xi_i=(1,0,0)/\sqrt{\gamma^{xx}}$, the fields ${}^\pm U$ read
\begin{eqnarray}
{}^\pm U &\equiv& 
 \frac{\tilde{A}^{xx}}{\gamma^{xx}} -\frac23e^{4\phi}K
\pm \sqrt{\frac53}\Big(
-\frac12 V^{xx}{}_k\frac{\gamma^{kx}}{(\gamma^{xx})^{\frac32}}\nonumber\\
&&
+\frac{7}{15}\frac{\tilde{\Gamma}^x}{\sqrt{\gamma^{xx}}}
-\frac43 e^{4\phi}Q_k\frac{\gamma^{kx}}{\sqrt{\gamma^{xx}}}\Big)
\end{eqnarray}
 
\noindent which can be inverted as
\begin{equation}
\tilde{A}^{xx}-\frac23\tilde{\gamma}^{xx}K
= \frac{\gamma^{xx}}{2}({}^- U +{}^+ U)
\end{equation}

\noindent Furthermore, since $\tilde{A}^{xx}-\frac23\tilde{\gamma}^{xx}K =
\tilde{\gamma}^{xj}(\tilde{A}^x_j-\frac23\delta^x_jK)$ then we have:
\begin{equation}
 \tilde{\gamma}^{xj}(\tilde{A}^x_j-\frac23\delta^x_jK) = 
\frac{\gamma^{xx}}{2}({}^- U +{}^+ U)
\end{equation}

\noindent Isolating the term with $j=x$ we have
\begin{equation}
 \tilde{\gamma}^{xx}(\tilde{A}^x_x-\frac23K) = 
\frac{\gamma^{xx}}{2}({}^- U +{}^+ U) 
- \tilde{\gamma}^{xy}\tilde{A}^x_y
- \tilde{\gamma}^{xz}\tilde{A}^x_z
\end{equation}

\noindent Using (\ref{AstoUs}) in the right-hand side, we finally have
\begin{eqnarray}
 \tilde{A}^x_x-\frac23K &= &
\frac{e^{4\phi}}{2}({}^- U +{}^+ U) 
-\frac{\tilde{\gamma}^{xy}}{2\tilde{\gamma}^{xx}}({}^- U_y + {}^+
U_y)\nonumber\\ 
&&
- \frac{\tilde{\gamma}^{xz}}{2\tilde{\gamma}^{xx}}({}^- U_z + {}^+ U_z)
\end{eqnarray}

\noindent Clearly, since the time derivatives of the incoming fields ${}^- U_y$
and ${}^- U_z$ are already prescribed by (\ref{Gxybssn}) and (\ref{Gxzbssn}),
then the boundary equation (\ref{Gxxbssn}) prescribes the time derivative of
the incoming field ${}^- U$ in terms of its outgoing counterpart. 

It only remains to understand the role of the fourth boundary equation
(\ref{Gxtbssn}).  The reader can verify that the combination 
\begin{equation}
{}^0U \equiv \tilde{\Gamma}^x-8\tilde{\gamma}^{xk}Q_k
\end{equation}

\noindent is one of the 18 characteristic fields with zero characteristic
speed. But this is exactly equal to the combination that appears under the
time-derivative sign in (\ref{Gxtbssn}) up to terms of lower order, that is
\begin{equation}
\tilde{\Gamma}^x-8\tilde{\gamma}^{xk}Q_k = \gamma^{xi}\big( (\ln(\det\gamma_{kl})),_{i}
    -\gamma^{kl}{\gamma}_{ik,l}\big)+\ldots
\end{equation}

\noindent Clearly, thus, the fourth boundary equation (\ref{Gxtbssn}) is a
condition on the time derivative of the ``static'' characteristic field
${}^0U$, and, as such, it should be identically satisfied.  The fourth equation
is not necessary in order to prescribe incoming fields.  

In summary, we have found that three of the six incoming characteristic fields
are prescribed by three of the four equations $G^x_a=0$.  These are precisely
the characteristic fields that travel at speeds different from the speed of
light. The fourth equation is redundant, but must be satisfied by a static
characteristic field as a consequence of the initial data and the evolution
equations.   

This is entirely consistent with the analysis of Section~\ref{sec:4}.  The
results of this Section show that, in the case of the first-order reduction,
there are only three boundary prescriptions in the set $G^x_a=0$, and the three
incoming fields that travel at the speed of light remain freely specifiable. 
This is very similar to the case of the Einstein-Christoffel formulation
discussed in \cite{boundary3d} and \cite{bcconst}. The fields that remain freely specifiable are labelled by the components
$\tilde{A}^y_z$ and $\tilde{A}^y_y - \tilde{A}^z_z$, and the trace $K$ of the
extrinsic curvature, as in the case of the Einstein-Christoffel formulation. The exception is that in the
present case it becomes clear that the incoming fields that remain free are
physical in the sense that they are the only ones traveling at the speed of
light (the same case cannot be made in the Einstein-Christoffel formulation
because all its incoming fields travel at light speed).

It is not clear to us whether any of the conclusions regarding the incoming
characteristic fields may be directly transferred from the first-order
reduction to the original BSSN equations, that is, Eqs.~(\ref{bssn}). The main
use of the results of this Section is to add insight into the boundary-value
problem of the (standard) BSSN equations, as well as to increase our
understanding of the role of the Einstein boundary conditions $G^x_a=0$ in the
initial-boundary-value problem of the Einstein equations across
formulations.

%-------------------------------------------------------------------
\section{Concluding remarks}\label{sec:6}

We have written down explicitly the principal terms of the four components of
the projection of the Einstein tensor perpendicularly to a timelike boundary in
terms of the variables of the BSSN formulation, Eqs.~(\ref{bssn}).  The
resulting expressions, Eqs.~(\ref{Gxybssn}), (\ref{Gxzbssn}), (\ref{Gxxbssn})
and (\ref{Gxtbssn}), contain no second derivatives across the boundary of any
of the fundamental variables, and no first derivatives across the boundary of
any of the three first-order variables $\tilde{\Gamma}^i$. Consequently,
setting them to zero results in four equations internal to the boundary in the
sense that they are constraints on the timelike boundary surface. Such boundary
equations can then be interpreted as boundary conditions for the BSSN
formulation of the Einstein equations.  Trivially, these boundary conditions
are necessary to the effect of satisfying the Einstein equations on the
boundary surface.  But far from trivially, we have demonstrated that three of
them are necessary in order to ensure that the solution of the BSSN equations
will satisfy the constraints in the future of the initial slice. 

A meaning to the three non-trivial boundary equations can be assigned as
follows.  A timelike boundary surface can be interpreted as an artificial
boundary separating two regions in a spacetime generated by the same Cauchy
surface (that is: a ``big'' spacetime with no boundary or with a null
boundary).  In this interpretation, the values on the timelike ``boundary''
surface are completely determined by evolution from the initial data. The
initial data satisfy the momentum and Hamiltonian constraints. Therefore, the
values on the timelike ``boundary'' surface must be related by some sort of
evolution of the constraints from their initial values.  Three of the
constraints propagate towards the ``boundary'' from data in the interior region
of the Cauchy surface.  But three of them propagate towards the ``boundary''
from Cauchy data in the other side of the timelike surface.  These three
``incoming'' constraints must affect the ``boundary'' values that are
interpreted as incoming, and a prescription for their effect on the boundary
values is needed.  The point is that the three nontrivial equations in the set
$G^x_a=0$ accomplish such a prescription by realizing, in a sense, the
evolution of the incoming initial constraints towards the ``boundary''
surface.    

Just like in the case of the ADM equations, the problem of the boundary
conditions of the BSSN formulation remains open in the following sense.  There
are, in principle, 15 variables whose boundary values are needed in order to
proceed forward in an integration in time. Of these, four can be obtained from
initial data by the four equations $G^x_a=0$, which are, in fact,
time-evolution equations internal to the boundary surface. It is not known, at
this time, how many of the remaining 11 variables are freely specifiable at the
boundary, and how many are determined by evolution from initial data. The
potentially freely specifiable variables represent an indeterminacy in the
sense that their role in shaping up the final solution is as important as that
of the initial data. Thus, physical insight would be required to prescribe the
(potentially free) boundary data in a manner analogous to the physical insight
required to determine the free initial data for a particular problem (e.g, for
binary black hole collisions). 

We have also studied the related problem of a strongly-hyperbolic reduction of
the BSSN equations, that is, Eqs.~(\ref{bssn1}).  In this case, the boundary
values that need to be specified are known to be those of six of the 30
fundamental variables, of which three are prescribed by the Einstein boundary
conditions. The others are determined, in principle, from the initial values
and the evolution equations. We have not concerned ourselves with the question
of determining which linear combinations of $G^x_a=0$ lead to a well-posed
initial-boundary-value problem, since our present first-order reduction is
meant only as an auxiliary device in order to shed light on the original
problem and is not expected to be used for numerical simulations. The problem
of separating the Einstein boundary conditions for this first-order reduction
into a well-posed set and an ill-posed set remains open at this time. 

Other issues that remain wide open are whether or not the (standard) BSSN
formulation with the Einstein boundary conditions can be implemented in a
numerically stable manner in the rigorous sense of finite-differences, and
whether the Einstein boundary conditions have any effect on the run time of
numerical simulations. Because the BSSN formulation is not strongly hyperbolic
in the standard sense, the former issue lies outside of our current interests. 
The latter issue is also of fundamental importance to numerical simulations
because it has been observed that constraint violations play a crucial role in
shortening the run time~\cite{teukolsky}. Our expectation is that the Einstein
boundary conditions, being ``constraint-preserving'' in the sense discussed in
Section~\ref{sec:4}, can potentially increase the run time of numerical
simulations, assuming that other potential numerical instabilities are under
control.

%--------------------------------------------------------------------
\begin{acknowledgments}
This work was supported by the NSF under grants No.
PHY-0244752 to Duquesne University, and No. PHY-0135390
to Carnegie Mellon University.
\end{acknowledgments}

%\bibliography{references}   % run latex, bibtex, latex, latex
%\bibliographystyle{prsty}

\end{document}